\documentclass[11pt,a4paper]{article}
\usepackage{makeidx}
\usepackage{graphicx}
\usepackage{latexsym}
\usepackage{amsmath}
\usepackage{amsfonts}
\usepackage{verbatim}
\usepackage{nameref}
\usepackage{hyperref}
\usepackage{sectsty}
\usepackage{enumerate}

\usepackage{amssymb}

\usepackage{amsmath}
 \usepackage[dvips]{epsfig}
\usepackage{amsfonts}
\usepackage{amsmath,amscd}
\usepackage{amsbsy}
\usepackage{amscd}
\usepackage{float}
\usepackage{rotating}
\usepackage{xy} \xyoption{all} \xyoption{import} \xyoption{rotate}
\usepackage{pst-plot}
\usepackage{psfrag}
\usepackage{a4wide}

\newcommand{\WW}{{W}}
\newcommand{\hatW}{\hat{W}}
\newcommand{\barW}{\bar{W}}
\newcommand{\alp}{\alpha}

\newcommand{\Oo}{\Omega}
\newcommand{\dO}{{\rm d} \Omega}
\newcommand{\dG}{{\rm d} \Gamma}

\newcommand{\Gt}{{\Gamma_t}}
\newcommand{\calB}{{\cal B}}
\newcommand{\uu}{{u}}

\newcommand{\xx}{{x}}

\newcommand{\half}{\frac{1}{2}}

\newcommand{\bB}{{\bf B}}

\newcommand{\bF}{{\bf F}}

\newcommand{\bn}{{\bf n}}

\newcommand{\bx}{{\bf x}}

\newcommand{\bchi}{\mbox{\boldmath$\chi$}}

\newcommand{\bsig}{{\bf S}} 
\newcommand{\bgamma}{\mbox{\boldmath$\gamma$}}
\newcommand{\btau}{{\mbox{\boldmath$\tau$}}}

\newcommand{\calP}{{\cal P}}

\newcommand{\calU}{{\cal{U}}}

\newcommand{\real}{\mathbb{R}}
\newcommand{\eb}{\begin{equation}}

\newcommand{\ee}{\end{equation}}

\def \div{\mbox{div\hskip 1pt}}

\title{Duality  in G. Saccomandi's Challenge on Analytical Solutions to Anti-plane Shear Problem  in Finite Elasticity}
 \author{David Yang  Gao    \\[0.2cm]
 \small   Federation University Australia, Mt Helen, VIC 3353, Australia  \\
 \small Research School of Engineering, Australian National University, Canberra, Australia}
\date{}
\begin{document}
\maketitle

\begin{abstract}
 This note is  a response to   G. Saccomandi's recent challenge
 by showing basic mistakes in his  conclusions.
 The proof is elementary, but leads to some fundamental results
 in correctly  understanding an extensively  studied problem in continuum mechanics.

\end{abstract}

\section{Problem  and Arguments}

The so-called anti-plane shear  deformation in the literature  \cite{knowles,knowles77} is simply defined by
\begin{equation}\label{defor}
\bchi  = \left\{ x_1, \;\;   x_2, \;\;
  x_3  + \uu(\xx_1, \xx_2) \right\}  : \Oo\subset \real^2  \rightarrow \calB \subset  \real^3.
\end{equation}
The only  displacement is  $\uu(x_1,x_2)$ in $x_3$ direction, which   is also the only unknown to be determined for any  given boundary value problems.
 The deformation gradient is
 \begin{equation}\label{1}
\mathbf{F}= \nabla \bchi =  \left\{ \frac{\partial \chi_i}{\partial x_j} \right\}  =
 \left(\begin{array}{ccc}
1 & 0 & 0 \\
0 &  1& 0 \\
\uu_{,1} & \uu_{,2} & 1 \end{array} \right)
\end{equation}
The principal invariants  of $\bB= \bF \bF^T$ are
   $I_1 = I_2 = 3 + \gamma^2, \;\;
 I_3 =   1 $,  where  $\gamma = |\bgamma|$ and $ \bgamma = \nabla \uu = \{ \uu_{,\alp} \} (\alp=1,2)$
 is the shear strain.
Thus,  the strain energy  $\WW(\bF)$  can be equivalently written in the following different forms
\eb\label{eq-ws}
\WW(\bF)  
= \barW(I_1,I_2) = \hatW(\bgamma).
\ee

Simply following the  local analysis adopted first by Knowles in \cite{knowles},
Saccomandi  claimed in \cite{sacc15} that the
anti-plane shear deformation  for   a homogeneous, isotropic, incompressible material
should be  governed by an over-determined system:
\eb\label{eq-dsig}
\div  \bsig(\uu, p)  = 0 \;\; \forall \bx \in \calB \subset \real^3,
\ee
where  $\bsig = \nabla \WW(\bF) $ is   the first Piola-Kirchhoff stress  defined by (Equation (10) in \cite{knowles77})
\eb
\bsig = 2\barW_1 \bF + 2 \barW_2 (I_1 \bF - \bB \bF) - p \bF^{-T}, \;\;  \barW_\alp = \partial \barW/\partial I_\alp, \;\; \alp = 1,2,
\ee
 ``while  $p= p(x_1,x_2,x_3)$ is an arbitrary function (the Lagrange multiplier associated with the incompressibility constraint), which must
 be of the form\footnote{See also Equation (2.22) in \cite{knowles}.}
 \eb\label{p}
 p = c x_3 + {\bar p}(x_1,x_2),
  \ee
  where $c$ is a constant" \cite{sacc15}.
Due to  three equations but two unknowns $(\uu, p)$, it was proved in \cite{knowles} that the strain energy must
 satisfy additional constraints. \\
{\bf Theorem (Knowles, 1976 \cite{knowles})} If
  the strain energy $\barW(I_1, I_2)$ is such that   the  ellipticity condition (i.e. the equation (3.5) in \cite{knowles} for $\lambda =1$)
  \eb\label{3.5}
   [2 \gamma (\barW_1 +  \barW_2)]_{, \gamma} > 0 \;\;
  \ee
 holds,  then
  the associated incompressible elastic material    admits nontrivial states of
anti-plane shear    if and only if $\barW(I_1, I_2)$ also satisfies
the following  constitutive constraint    (i.e. equation (3.22) in \cite{knowles} for $\lambda =1$)
  \eb\label{3.22}
  b \barW_1 + (b   - 1) \barW_2  = 0   \mbox{ for some constant } \; b , \;\;\forall I_1= I_2  = 3  + \gamma^2, \;\;  \gamma = |\bgamma| \ge 0.
  \ee

 As highly cited papers  \cite{knowles,knowles77},  Knowles' over-determined system has been extensively
 applied to many anti-plane shear deformation problems   in literature, see
 G. Saccomandi's  recent papers \cite{sacc15,pucci-r-s14,pucci-sacc,pucci-sacc1} and references cited therein.\\

 Dual to the local analysis on the strong form of the equilibrium problem (\ref{eq-dsig}),
 Gao's approach   \cite{gao-cmt15} is based on
 minimum total potential principle, i.e.
 \eb
 (\calP): \;\; \min\left\{ \Pi(\uu) = \int_\Oo \hatW(\nabla \uu) \dO - \int_\Gt \uu t \dG \;\; |
 \;\; \uu \in \calU_c \right\}
 \ee
 where  $t: \Gt \subset \partial \Oo \rightarrow \real$ is  a given boundary shear force,
 $\calU_c$ is the {\em kinetically admissible space}, in which, homogeneous boundary condition is given.
Since the  incompressible condition $\det \bF(\uu) \equiv 1$ is trivially satisfied,
$\calU_c$ is a convex set, the weak variation  $\delta \Pi(\uu) = 0$  leads to only one
equilibrium equation in $x_3$ direction:
\eb\label{eq-hw}
\div \btau (\nabla\uu) = 0 \;\; \mbox{ in } \Oo, \;\; \bn \cdot  \btau (\uu)  = t \;\; \mbox{ on } \Gt,
\ee
where $\btau(\bgamma) = \nabla  \hatW(\bgamma)$ is the shear stress, $\bn = \{ n_\alp\}$
is a unit norm vector on $\partial \Oo$.
The author proved  in \cite{gao-cmt15} that
for any given $t(\bx) \neq 0 $ on $ \Gt$, this well-defined fully nonlinear PDE has at   least one nontrivial  solution $\uu$, which  can be obtained analytically
by a canonical duality theory developed in \cite{gao-dual00}.
Additionally, both global minimizer and local extrema (i.e. local min and local max) can be identified by
an  associated triality theory \cite{gao-cmt15}.\\

In G. Saccomandi's recent paper \cite{sacc15},
instead of finding directly any possible mistake  in \cite{gao-cmt15},
 his arguments are based on   Knowles's   over-determined system. He claimed
 ``a major mistake contained in the paper \cite{gao-cmt15}"   due to the following  two issues in his  conclusions:

  ``The results  contained in \cite{gao-cmt15} are only valid

  $\bullet$ for a very special class of unconstrained (or compressible) materials;

    $\bullet$ for a very special class of incompressible materials and in the special case that the gradient of the pressure in the $x_3$ direction is null (i.e., $c = 0$)."

\section{Duality in Arguments}
It is  interesting to see  a  multi-level duality pattern  in  the arguments:

{\bf Level 1}:   By using opposite approaches (local vs. global),
 Knowles and Gao obtained two different systems (over-determined vs. determined).

 {\bf Level 2}:  If  
 G. Saccomandi's   conclusions in \cite{sacc15} hold, then his related works are correct,
 but  Gao's paper contained ``a major mistake". 
Dually, if Gao is correct, then  
Saccomandi's both conclusions  and his related works (such as \cite{pucci-sacc}) must be   wrong.\\

The prove of the dual statement in Level 2 is  elementary.

First, by the facts that for any given anti-plane shear deformation problems,
the incompressibility condition $\det \bF(\uu) \equiv 1$ is trivially
satisfied and the equivalent forms (\ref{eq-ws}) hold
without any additional constitutive constraints,
  any  critical solution to the variational problem $(\calP)$ must satisfy the  condition  $\det \bF(\uu) \equiv 1$.
  Also, the author never claimed that the results in \cite{gao-cmt15} are valid for general constrained materials.
    Therefore, Saccomandi's first conclusion is incorrect.

  Second, by the fact that the strain energy $\barW(I_1,I_2)$ depends only on $(x_1,x_2) \in \Oo$, the
  stress $\bsig$ is independent of $x_3$. Also, since the condition $\det \bF =1$ is independent of 
   $x_3$, its Lagrange multiplier $p$ should  be independent of $x_3$  
   and  we must have  $c \equiv 0$ in (\ref{p}).
  Therefore, the  Saccomandi's second  conclusion is wrong.

 Indeed,  even Knowles himself found in his second paper
 that ``all quantities are independent of $x_3$" (the first line on page 4 \cite{knowles77}).
Also, in \cite{sacc15} Saccomandi  let $p = - 2 W_3(I_1,I_2,1)$. If he is careful, he should know
that the gradient of the pressure in the $x_3$  direction must be  null.
Unfortunately, this obvious mistake happened in his related works, see page 168 in \cite{pucci-sacc}.

 It is well-known that the   equilibrium equations obtained (under certain regularity conditions)  by
  potential variational  principle are naturally compatible regardless of any possible constitutive laws.
  Since  $\det \bF(\uu) \equiv 1$ is trivially satisfied, the incompressible condition is not a variational constraint.
  According  to    the complementarity condition $p (\det \bF - 1) = 0$ in  KKT theory  \cite{lat-gao-ol15},
 the Lagrange multiplier  $p $ in \cite{sacc15}  could be any arbitrary nonzero  parameter, but can't be considered as an unknown variable.
 By the fact that
 \eb\label{ellip}
 \btau = \nabla  \hatW(\bgamma)  =
 \frac{\partial \barW(I_1,I_2) }{\partial \bgamma} = \barW_1 \frac{\partial I_1}{\partial \bgamma}
 +  \barW_2 \frac{\partial I_2}{\partial \bgamma} = 2 \bgamma [\barW_1 + \barW_2]
 \ee
 Saccomandi's  equation (1.7) in \cite{sacc15} is exactly the same as
 Gao's equation  (\ref{eq-hw})  in $x_3$ direction.
 Since there is no displacement  in both  $x_1$ and $x_2$ directions,  Saccomandi's two extra equilibrium equations  (1.6)  in   \cite{sacc15}   can't be obtained via either potential variational principle or
 the virtual-work principle. Therefore, these two equations are not needed for the problem considered.
 Indeed, due to the  arbitrary Lagrange multiplier $p$ in these two equations,
they  could be self-balanced, but are useless for the anti-plane shear deformation problem.

 Finally, let us check the constitutive constraints Saccomandi emphasized in his challenge.
 By the condition $I_1=I_2$ and chain rule in calculus, we must have $\barW_1= \barW_2 (\partial I_2/\partial I_1 )= \barW_2$. Thus, Knowles' constitutive constraint
 (\ref{3.22}) is naturally satisfied for $b= 1/2$.
 Also by (\ref{ellip})  one  can easily check that Knowles' ellipticity condition  (\ref{3.5})
 is  just a special case of the strong Legendre condition $\nabla^2 \hatW(\bgamma) \succ 0 \;\; \forall \bgamma$,
 which can only guarantee the convexity of $\WW(\bF) = \hatW(\bgamma)$.
Therefore, Knowles' constitutive constraints are neither necessary nor sufficient for
 an incompressible homogeneous  elastic material  to   admit  nontrivial states of
anti-plane shear.
Indeed, by the canonical duality theory,
complete sets of analytical solutions have been obtained for general nonconvex finite deformation problems,
and the existence of these solutions depends   mainly on   boundary conditions and external forces
\cite{gao-cmt15,gao-haj,gao-ogden-zamp,gao-ogden-qjmam}.

 Ellipticity in PDEs   is a classical concept originally from linear systems where the stored energy is a quadratic function  $ \hatW(\nabla \uu) = \half  H_{\alp\beta} \uu_{,\alp} \uu_{,\beta} $  and the linear
 operator $L[\uu] = - [H_{\alp\beta} \uu_{,\beta} ]_{,\alp}$ is elliptic   if the Hessian
 matrix  $\{H_{\alp\beta}\} \succ 0$.
 However, this definition is only for convex systems. It is shown in
  the recent paper   \cite{antichal} that for nonconvex systems, the ellipticity depends not only on the
  stored energy $\WW(\bF)$, but also on  the external force field.
  For any given nonconvex $\WW(\bF)$, the problem can have unique solution as long as the external force is bigger
  enough. This result is naturally included in the triality theory with extensive applications in
  multidisciplinary fields of nonconvex analysis and global optimization
    \cite{gao-bridge}.

 \section{Conclusions}

The  conclusions contained in \cite{sacc15} are wrong. The proof is truly elementary.

 This mistake leads to many other  problems in G. Saccomandi's related papers.
 By the fact that $I_1=I_2, \;\; I_3=1$,
the  anti-plane shear  deformation of a homogeneous elastic body
must be governed by a generalized neo-Hookean model,   i.e.    $\WW(\bF) = \barW(I_1)$.
The proof is trivial   \cite{antichal}.

The  constitutive constraint  in \cite{knowles} has been reconsidered, which is automatically satisfied for $b = \lambda/2$.
The ellipticity condition in \cite{knowles,knowles77}
is  neither necessary nor sufficient for an isotropic
 homogeneous elastic material to admit nontrivial
states of anti-plane shear deformation.
For nonconvex systems,  the ellipticity of a fully nonlinear boundary value problem  depends
not only on the stored energy, but also on the external force. The triality theory can be used to identify both
global minimizer and local extrema.
Detailed study is given in recent paper \cite{antichal}.

The results presented in \cite{gao-cmt15} are valid for general nonconvex anti-plane shear deformation problems as long as the equivalent form  (\ref{eq-ws}) holds for the strain energy functions $\barW$ and $\hatW$.
Also Gao's analytical  solutions are obtained from global analysis in Banach space,  the nonsmoothness is allowed.
Dually, even if the overdetermined problem addressed in  Saccomandi's papers is correct,
which allows only unique smooth solution  due to regularity and ellipticity constraints.
By the fact that the mistakes in \cite{sacc15} are elementary and also contained  in \cite{pucci-sacc}, readers must be careful with entire results  presented in \cite{pucci-r-s14,pucci-sacc,pucci-sacc1}.


\begin{thebibliography}{99}






\bibitem{knowles} Knowles, J.K. (1976).
On finite anti-plane shear for imcompressible elastic materials,
{\em J. Australia Math. Soc.}, 19, 400-415.

  \bibitem{sacc15} Saccomandi G. (2015). D. Y. Gao: Analytical solutions to general anti-plane shear problems in finite elasticity, {\em Continuum Mech. Thermodyn.}


\bibitem {gao-cmt15}Gao, D.Y. (2015)
Analytical solutions to general anti-plane shear problem in finite elasticity.
{\em Continuum Mech Theorm. }, 2015.


\bibitem{gao-dual00}  Gao, D.Y. (2000). \emph{Duality Principles in Nonconvex
Systems: Theory, Methods and Applications}, Kluwer Academic
Publishers,  Dordrecht /Boston /London, { xviii + 454pp. }



\bibitem{gao-haj}Gao, DY and Hajilarov, E. (2015). Analytic solutions to three-dimensional
finite deformation problems governed by
St Venant–Kirchhoff material, {\em Math Mech Solids}, DOI: 10.1177/1081286515591084

\bibitem{gao-ogden-zamp} Gao, D.Y. and Ogden, R.W. (2008).
Closed-form solutions, extremality and nonsmoothness criteria in a large deformation elasticity problem,
{\em ZAMP,} 59:498 - 517.

\bibitem{gao-ogden-qjmam} Gao, D.Y. and Ogden, R.W. (2008).
 Multiple solutions to non-convex variational problems with implications for phase transitions and numerical computation,  {\em Quarterly J. Mech. Appl. Math. }  61 (4), 497-522.

\bibitem{gao-bridge} Gao, DY, Ruan, N, and Latorre, V (2015).
Canonical duality-triality: Bridge between nonconvex analysis/mechanics and global optimization in complex systems.
{\em Math. Mech. Solids}.


\bibitem{antichal} Gao, DY (2015).
Remarks on analytical solutions in nonlinear elasticity and anti-plane shear   problem,
 accepted, http://arxiv.org/abs/1507.08748



\bibitem{knowles77}Knowles, J. K. (1977).
 On note on anti-plane shear for compressible materials in finite elastostatics.
 {\em Journal of Australian Mathematical Society B,} 20, 1–7.




 \bibitem{lat-gao-ol15} Latorre, V. and Gao, D.Y. (2015).
 Canonical duality for solving general nonconvex
constrained problems, {\em Optimization Letters}, DOI 10.1007/s11590-015-0860-0



\bibitem{pucci-r-s14}Pucci, E., Rajagopal, K.R., and Saccomandi, G. (2014).
 On the determination of semi-inverse solutions of nonlinear
 Cauchy elasticity: The not so simple case of anti-plane shear. {\em Int. J. Engineering Sciences}. http://dx.doi.org/10.1016/j.ijengsci.2014.02.033

\bibitem{pucci-sacc} Pucci E., Saccomandi G. (2013).
The anti-plane shear problem in non-linear elasticity revisited.
{\em  Journal of Elasticity},  113, 167-177.

\bibitem{pucci-sacc1} Pucci E., Saccomandi G. (2013). Secondary motions associated with
anti-plane shear in nonlinear isotropic elasticity, {\em Q. Jl Mech. Appl. Math,} Vol. 66. No. 2, 221-239.



\end{thebibliography}
\end{document}